\documentclass[showpacs,preprintnumbers,amsmath,amssymb]{revtex4}
\newtheorem{theorem}{Theorem}

\newtheorem{lemma}{Lemma}

\usepackage{graphicx}
\usepackage{dcolumn}
\usepackage{bm}

\begin{document}
\title{
On Ping-Pong protocol and its variant
}

\author{Takayuki Miyadera$\ ^{1}$}
\email{miyadera-takayuki@aist.go.jp}
\author{Masakazu Yoshida$\ ^{2}$}
\author{Hideki Imai$\ ^{3}$}%
\affiliation{%
$\ ^{1,3}$
Research Center for Information Security (RCIS), \\
National Institute of Advanced Industrial
Science and Technology (AIST). \\
Daibiru building 1003,
Sotokanda, Chiyoda-ku, Tokyo, 101-0021, Japan.
}%
\affiliation{
$\ ^{2,3}$
Graduate School of Science and Engineering,
\\
Chuo University.
\\
1-13-27 Kasuga, Bunkyo-ku, Tokyo 112-8551, Japan .
 }%


\date{\today}

\begin{abstract}
We discuss the Ping-Pong protocol which was proposed by 
Bostr\"om and Felbinger. 
We derive a simple trade-off inequality between 
distinguishability of messages for Eve and detectability of Eve for
legitimate users. 
Our inequality holds for arbitrary initial 
states. 
That is, even if Eve prepares an initial state, she cannot
distinguish messages without being detected. 
We show that the 
same inequality holds also on another protocol in which Alice and Bob 
use one-way quantum communication channel twice. 
\end{abstract}
\maketitle
\section{Introduction}
In 2002, Bostr\"om and Felbinger \cite{BF02} 
proposed a quantum protocol
which is called
Ping-Pong protocol. 
Being different from other protocols 
such as BB84 or E91, this protocol 
uses two-way quantum communication. 
They showed a trade-off inequality between information gain by Eve 
and the error probability detected by Alice and Bob 
on the ideal setting of the protocol. 
That is, information gain by Eve is inevitably detected 
by Alice and Bob. 
While they insist that this protocol works 
as a secure direct communication as well as a key distribution protocol, 
there have been several discussions on its 
security from various points of view
\cite{QCai03,Woj03,QCai06,BF08}. 
The purpose of the present 
paper is not to discuss the security issue of the protocol 
but to give a simple derivation of another trade-off inequality 
between distinguishability of messages for Eve and detectability of Eve for
legitimate users.
The inequality holds for arbitrary initial states. Thus even if 
an initial state is prepared by Eve, she cannot distinguish 
the messages without being detected. As a byproduct, we 
show that the same inequality holds on a variant of the 
original protocol in which Bob sends his quantum system twice to Alice. 
\par
This paper is organized as follows. In the next section, 
we give a short description of the original Ping-Pong protocol.
In section \ref{analysis}, a 
trade-off inequality is derived in a simple manner. 
In section \ref{variant}, a variant of the original protocol 
is given. It is shown that our trade-off inequality still holds 
on this variant. 
\section{Protocol}
In this section we give a brief explanation on the simplest 
version of the protocols for 
Alice to send Bob one-bit message (or secret key). 
Bob first prepares a maximally entangled state 
$|\phi_0\rangle :=\frac{1}{\sqrt{2}}(|11\rangle +|00\rangle)$. 
He sends one of the bipartite systems which is called 
system A. It is described by a Hilbert space
${\cal H}_A (\simeq {\bf C}^2)$. 
Another system possessed by Bob is 
called system B with its Hilbert space 
${\cal H}_B (\simeq {\bf C}^2)$. 
Bob confirms Alice's receipt of the system A \cite{confirm}. 
Alice randomly chooses one from $\{\mbox{Control},\ \mbox{Message}\}$. 
If she chose ``Control", she lets Bob know it and 
they both make measurements of $\sigma_z(A)$ and $\sigma_z(B)$ 
on their own systems respectively \cite{sigma}. 
If their outcomes disagree, they know existence of Eve and abort the protocol. 
On the other hand, if Alice chose ``Message", she encodes her 
one-bit message to her system A. She does nothing on 
system A for the message $0$. She operates $\sigma_z(A)$ 
on it for the message $1$, which changes the phase 
with respect to $|0\rangle$. 
Alice sends back the system A to Bob. 
Bob makes a Bell measurement on the composite system A and B
to know the encoded message. 
As pointed out in \cite{QCai03,BF08}, this naive protocol yields a simple 
``attack" that disturbs the message without being detected. 
That is, just an attack only on the second quantum communication 
from Alice to Bob does not affect the error probability 
in the control mode but can change the message while Eve cannot 
obtain any information. As claimed 
in \cite{QCai03,BF08}, this 
disadvantage may be avoided by introducing authentication 
phase after the protocol or slightly modifying the protocol 
itself. We, however, do not treat this problem here. 
What we are interested in is whether Eve can distinguish 
the messages $0$ and $1$ without being detected. 
\section{Analysis}\label{analysis}
Let us see what Eve can do in this protocol. 
Eve prepares her own system E which 
is described by a Hilbert space ${\cal H}_E$. 
We write the initial state of system E as $|\Omega\rangle$.  
She interacts 
it with system A when system A is sent between Alice and Bob. 
That is, she has two chances to obtain the information.
Let us denote the first interaction by a unitary map
$W :{\cal H}_A  \otimes {\cal H}_E \to {\cal H}_A \otimes {\cal H}_E$
and
the second interaction by 
$V: {\cal H}_A \otimes {\cal H}_E \to {\cal H}_A \otimes {\cal H}_E$. 
The state after the first attack is described by 
$|\Psi\rangle:=W |\phi \otimes \Omega\rangle$. 
The final state over the tripartite system A, B and E 
in a message mode becomes 
$V|\Psi\rangle$
when Alice's message is $0$ and 
becomes 
$V\sigma_z(A)|\Psi\rangle$
when Alice's 
message is $1$. Eve's purpose is to distinguish them. 
The states to be distinguished by Eve are
\begin{eqnarray*}
\rho_0&:=& \mbox{tr}_{AB}\left(V|\Psi \rangle
\langle \Psi| V^* \right) 
\\
\rho_1&:=& \mbox{tr}_{AB}\left(V\sigma_z(A)|\Psi \rangle
\langle \Psi | \sigma_z(A) V^*
\right) . 
\end{eqnarray*}
We employ fidelity \cite{Uhlmann,Jozsa}
as a measure for (in)distinguishability of states. 
The fidelity between two states $\rho$ and $\sigma$ is 
defined by $F(\rho,\sigma):=\mbox{tr}\sqrt{\rho^{1/2} \sigma \rho^{1/2}}$. 
It takes $1$ if and only if $\rho=\sigma$ and takes a
nonnegative value less than $1$ in general. 
The key lemma is the following which
played an important role in \cite{MIWig}
to derive a version of Wigner-Araki-Yanase theorem. 
\begin{lemma}\label{lemma1}
Suppose that we have two systems that are described 
by Hilbert spaces ${\cal H}_1$ and ${\cal H}_2$, and 
a pair of pure states $|\phi_0 \rangle, |\phi_1 \rangle \in {\cal H}_1
\otimes {\cal H}_2$.  
If we put states on ${\cal H}_2$ as
\begin{eqnarray*}
\rho_j :=\mbox{tr}_1 \left(|\phi_j \rangle
\langle \phi_j |
 \right),
\end{eqnarray*}
for $j=0,1$, then for an arbitrary operator $X$ on ${\cal H}_1$, 
\begin{eqnarray*}
|\langle \phi_0 |X|\phi_1 \rangle |
\leq \Vert X\Vert F(\rho_0,\rho_1)
\end{eqnarray*}
holds, where $\Vert \cdot\Vert$ is an operator norm 
defined by $\Vert X\Vert:=\sup_{|\phi\rangle \neq 0}
\frac{\Vert X|\phi\rangle \Vert}{\Vert |\phi\rangle\Vert}$.
\end{lemma}
{\bf Proof:}
\\
We consider an arbitrary positive-operator-valued measure 
(POVM) $\{E_{\alpha}\}$ 
on ${\cal H}_2$, that is, every 
positive operator $E_{\alpha}$ acts only on ${\cal H}_2$ and 
satisfies $\sum_{\alpha}E_{\alpha}={\bf 1}$. 
We obtain
\begin{eqnarray*}
|\langle \phi_0|X|\phi_1\rangle |
=|\sum_{\alpha} \langle \phi_0| E_{\alpha} X|\phi_1\rangle |
=|\sum_{\alpha} \langle \phi_0| E_{\alpha}^{1/2} X E_{\alpha}^{1/2} |
\phi_1 \rangle |,
\end{eqnarray*} 
where we used the commutativity between
 $E_{\alpha}^{1/2}$ and $X$. 
 We further obtain
\begin{eqnarray*}
|\langle \phi_0|X|\phi_1\rangle | 
&\leq & \sum_{\alpha}|\langle \phi_0|E_{\alpha}^{1/2}
X E_{\alpha}^{1/2} |\phi_1\rangle | \\
&\leq& \sum_{\alpha} \langle \phi_0 |E_{\alpha} |\phi_0 \rangle^{1/2}
\langle \phi_1 |E_{\alpha}^{1/2} X^* X E_{\alpha}^{1/2} |\phi_1
\rangle^{1/2} \\
&\leq  & \sum_{\alpha} \langle \phi_0 |E_{\alpha} |\phi_0 \rangle^{1/2}
\langle \phi_1 |E_{\alpha} |\phi_1
\rangle^{1/2} \Vert X\Vert,
\end{eqnarray*}
where we used the Cauchy-Schwarz inequality to derive 
the second line and the definition of the operator norm 
to derive the third line. 
By using a property $F(\rho,\sigma)=
\inf_{E:POVM} \sum_{\alpha}
\sqrt{
\mbox{tr}(\rho E_{\alpha})\mbox{tr}(\sigma E_{\alpha})}$
which was shown in \cite{FC95,BCFJS}, 
we take the infimum of the above inequality 
over all the POVMs to obtain
\begin{eqnarray*}
|\langle \phi_0|X|\phi_1\rangle | 
\leq \Vert X \Vert F(\rho_0, \rho_1).
\end{eqnarray*}
It ends the proof.
\hfill {\bf Q.E.D.}
\par
In applying this lemma to Wigner-Araki-Yanase theorem, 
it was important to have a conserved quantity. 
Also in the Ping-Pong protocol, we have a conserved quantity. 
In fact, since the system B is kept by Bob during whole the 
protocol, the attack does not give any effect on 
the operator on system B. That is, for any operator 
$X$ on ${\cal H}_B$, 
$W^* V^* X V W =X$ and $V^* XV=X$ hold. We take the second equation 
and operate $\langle \Psi| $ 
and $\sigma_z(A)|\Psi \rangle$ to it. 
We obtain
\begin{eqnarray*}
\langle \Psi | V^*X V \sigma_z(A)|\Psi
\rangle
&=&\langle \Psi | X \sigma_z(A)| \Psi
\rangle .  
\end{eqnarray*}
Taking the absolute value of the above equation, 
we apply the lemma with 
${\cal H}_1={\cal H}_A\otimes {\cal H}_B$, 
${\cal H}_2={\cal H}_E$,
$|\phi_0\rangle=V |\Psi \rangle$ 
and $|\phi_1 \rangle=V \sigma_z(A) |\Psi \rangle$ 
to obtain, 
\begin{eqnarray*}
\Vert X \Vert F\left(\rho_0,\rho_1
\right)
\geq
|\langle \Psi| X  \sigma_z(A)|\Psi
\rangle |.
\end{eqnarray*}
Thus the indistinguishability 
 of the messages for Eve is bounded from below 
 by a correlation function after the first attack. 
 If we put $X=\sigma_z(B)$, this correlation function becomes
 \begin{eqnarray*}
 \langle \Psi| \sigma_z(B) \sigma_z(A)
 |\Psi \rangle
 =p(0,0)+p(1,1)-p(0,1)-p(1,0)=1-2p(\sigma_z(A)\neq \sigma_z(B)),
 \end{eqnarray*}
 where $p(i,j)$ is probability for Alice and Bob to obtain 
 $\sigma_z(A)=i$ and $\sigma_z(B)=j$ respectively in $|\Psi\rangle$. 
 That is, this is a probability distribution of the 
 outcomes in the control mode. Thus we obtain
 \begin{eqnarray*}
 \left| 1-2p(\sigma_z(A)\neq \sigma_z(B))
 \right|
 \leq F(\rho_0,\rho_1).
 \end{eqnarray*}
 Note that this inequality holds for an arbitrary 
 state $|\Psi\rangle$ over the tripartite state
 since we did not use its concrete form. 
Thus we proved the following theorem. 
\begin{theorem}
In the Ping-Pong protocol, Eve cannot distinguish the messages
$0$ and $1$ without being detected. In fact, if we put 
$p(\sigma_z(A)\neq \sigma_z(B))$ probability for Alice 
and Bob to obtain 
different outcomes in the control mode, 
indistinguishability measured by the fidelity is 
bounded as
\begin{eqnarray*}
\left| 1-2p(\sigma_z(A)\neq \sigma_z(B))\right|
\leq F(\rho_0,\rho_1). 
\end{eqnarray*}
Here the initial state can be arbitrary.
Even if it was prepared by Eve, the above trade-off 
inequality still holds. 
\end{theorem}
It should be remarked that although the above trade-off 
inequality holds for arbitrarily prepared states, 
it does not mean that the protocol works in 
such cases. In fact, in such cases Alice and Bob 
cannot share the messages even if they do not 
detect Eve. That is, success in 
message sharing and information gain by Eve are
different matters in this protocol.
\section{A variant of the protocol}\label{variant}
In the original Ping-Pong protocol Bob first sends a 
qubit to Alice and receives it in the end of the protocol. 
In this section, we consider its variant.
After the confirmation of Alice's receipt of a qubit, 
Bob, instead of Alice, sends a message to Alice. 
For the definiteness, we describe the whole protocol in the following. 
Bob first prepares a maximally entangled state 
$|\phi_0\rangle :=\frac{1}{\sqrt{2}}(|11\rangle +|00\rangle)$. 
He sends one of the bipartite systems which is called 
system A. It is described by a Hilbert space
${\cal H}_A$. 
Another system possessed by Bob is 
called system B with its Hilbert space 
${\cal H}_B$. Bob confirms Alice's receipt of the system A. 
Bob randomly chooses one from $\{\mbox{Control},\ \mbox{Message}\}$. 
If he chose ``Control", he lets Alice know it and 
they both make measurements of $\sigma_z(A)$ and $\sigma_z(B)$ 
on their own systems respectively. 
If their outcomes disagree, they know existence of Eve and abort the protocol. 
On the other hand, if Bob chose ``Message", he encodes his 
one-bit message to his system B. He does nothing on 
system B for the message $0$. He operates $\sigma_z(B)$ 
on it for the message $1$, which changes the phase 
with respect to $|0\rangle$. 
Bob sends the system B to Alice. 
Alice makes a Bell measurement to the composite system A and B
to know the encoded message. 
\par
We can prove again the following theorem. 
\begin{theorem}
In the above variant of the Ping-Pong protocol, 
Eve cannot distinguish the message
$0$ and $1$ without being detected.
Let us denote by $\mu_0$ Eve's final state 
corresponding to the message $0$ and $\mu_1$ 
one corresponding to the message $1$. 
If we put 
$p(\sigma_z(A)\neq \sigma_z(B))$ probability for Alice 
and Bob to obtain 
different outcomes in the control mode, 
indistinguishability between 
$\mu_0$ and $\mu_1$ is 
bounded as
\begin{eqnarray*}
\left| 1-2p(\sigma_z(A)\neq \sigma_z(B))\right|
\leq F(\mu_0,\mu_1). 
\end{eqnarray*}
Here the initial state can be arbitrary.
Even if it was prepared by Eve, the above trade-off 
inequality still holds. 
\end{theorem}
{\bf Proof:}
\par
The proof runs in the same manner with the previous theorem.
Eve, with her own system E, interacts system A and system B 
when they are sent from Bob to Alice. 
We denote by $|\Psi\rangle$ the state over system 
A, B and E after the 
first attack
and denote the second attack by a unitary map
$U: {\cal H}_B \otimes {\cal H}_E
\to {\cal H}_B \otimes {\cal H}_E$. 
In the message mode, the states Eve wants to 
distinguish are $\mu_0:=\mbox{tr}_{AB}(U|\Psi\rangle
\langle \Psi |U^*)$ and $\mu_1:=\mbox{tr}_{AB}
(U \sigma_z(B)|\Psi\rangle \langle \Psi|\sigma_z(B)U^*)$
Since the second attack does not change the operator on 
${\cal H}_A$, $U^* \sigma_z(A) U =\sigma_z(A)$ holds. 
We operate $\langle \Psi|\cdot \sigma_z(A)|\Psi\rangle$ 
on this equation to obtain, 
\begin{eqnarray*}
\langle \Psi|U^* \sigma_z(A) U \sigma_z(B)|\Psi\rangle
=\langle \Psi|\sigma_z(A) \sigma_z(B)|\Psi\rangle.
\end{eqnarray*} 
Applying Lemma \ref{lemma1} to the absolute value of the left hand side
with ${\cal H}_1={\cal H}_A\otimes {\cal H}_B$, 
${\cal H}_2 ={\cal H}_E$, $X=\sigma_z(A)$, $|\phi_0\rangle
=U|\Psi\rangle$ and $|\phi_1\rangle=U\sigma_z(B)|\Psi\rangle$,
we obtain, 
\begin{eqnarray*}
\left| 1-2p(\sigma_z(A)\neq \sigma_z(B))\right| \leq F(\mu_0,\mu_1).
\end{eqnarray*}
It ends the proof. 
\hfill Q.E.D.
\section{discussions}
In this paper, we treated 
the Ping-Pong protocol and 
derived a trade-off inequality between distinguishability of 
states for Eve and detectability for legitimate users. 
The inequality holds for arbitrary states that may be 
prepared even by Eve. We showed that the same inequality holds 
in a slightly different protocol 
in which the quantum communication is one-way. 
It, however, should be remarked that this trade-off relation 
does not directly mean the security of the protocols. 
For instance, Eve can change the message without being detected 
by making an attack only on the second communication phase. 
Furthermore, if Alice and Bob intend to use the protocols 
for direct communication, they need to confirm sufficiently 
many times the 
cleanness of the line before sending a message. 
In fact, otherwise Eve may obtain the message
with non-negligible probability. Thus further 
investigation on definition and analysis of the security should be needed. 


\begin{thebibliography}{9}
\bibitem{confirm}
Alice, for instance, can confirm her receipt of a particle 
by measuring number operator which does not destroy 
internal degrees of freedom. She then announces the receipt to Bob.
\end{thebibliography}
\end{document}